\documentclass{article}

\usepackage[nonatbib,preprint]{neurips_2022}

\usepackage[utf8]{inputenc} 
\usepackage[T1]{fontenc}    
\usepackage{hyperref}       
\usepackage{url}            
\usepackage{booktabs}       
\usepackage{amsfonts}       
\usepackage{nicefrac}       
\usepackage{microtype}      
\usepackage{xcolor}         
\usepackage{graphicx}
\usepackage{subfig}
\usepackage{wrapfig}
\usepackage{makecell}
\usepackage{amsmath}
\usepackage{amssymb}
\usepackage[natbib=true,maxnames=6,mincitenames=1,maxcitenames=1,style=ieee,url=false,eprint=false,isbn=false]{biblatex}
\AtEveryBibitem{\clearlist{language}}

\title{Differentiable programming for functional connectomics}

\author{%
  Rastko~Ciric \\
  Department of Bioengineering \\
  Stanford University\\
  Stanford, CA 94041 \\
  \texttt{rastko@stanford.edu} \\
  \And
  Armin W.~Thomas \\
  Stanford Data Science \\
  Stanford University \\
  Stanford, CA 94041 \\
  \And
  Oscar~Esteban \\
  Department of Radiology \\
  Universit\'e de Lausanne \\
  Lausanne, Switzerland \\
  \And
  Russell A.~Poldrack \\
  Department of Psychology \\
  Stanford University \\
  Stanford, CA 94041 \\
}

\begin{document}

\maketitle

\begin{abstract}
  Mapping the functional connectome has the potential to uncover key insights into brain organisation.
  However, existing workflows for functional connectomics are limited in their adaptability to new data, and principled workflow design is a challenging combinatorial problem.
  We introduce a new analytic paradigm and software toolbox that implements common operations used in functional connectomics as fully differentiable processing blocks.
  Under this paradigm, workflow configurations exist as reparameterisations of a differentiable functional that interpolates them.
  The differentiable program that we envision occupies a niche midway between traditional pipelines and end-to-end neural networks, combining the glass-box tractability and domain knowledge of the former with the amenability to optimisation of the latter.
  In this preliminary work, we provide a proof of concept for differentiable connectomics, demonstrating the capacity of our processing blocks both to recapitulate canonical knowledge in neuroscience and to make new discoveries in an unsupervised setting.
  Our differentiable modules are competitive with state-of-the-art methods in problem domains including functional parcellation, denoising, and covariance modelling.
  Taken together, our results and software demonstrate the promise of differentiable programming for functional connectomics.
\end{abstract}

\section{Introduction}

Many scientific disciplines depend on complex analytic workflows to extract salient information from data. In the life sciences, and in large-scale brain mapping in particular, the development and refinement of these workflows has grown to become an informatic subdiscipline in its own right. Although the introduction of better tools is a necessary vector of progress, informatic practice also comes at the cost of increased analytic flexibility, or ``researcher degrees of freedom''  \citep{carp_plurality_2012,Botvinik-Nezer2020-us}. In other words, the development of new instruments presents researchers with the combinatorial challenge of selecting, from among the growing set of available informatic tools, an analytic workflow conditioned on their dataset and scientific question. This challenge is significant for two reasons. First, many reported results are sensitive to particular pipeline configurations and fail to replicate under alternative configurations \citep{carp_plurality_2012,Botvinik-Nezer2020-us}. Second, publication bias obscures the often iterative process of tool development, and a skewed scientific incentive structure motivates informaticians to refine algorithms until they outperform state-of-the-art (SOTA) methods on benchmark datasets. Anecdotally, this is often accompanied by either improper or incomplete implementation of the SOTA baselines, or insufficient consideration of benchmark dataset properties that might lead to inflated performance.

The problem of designing a data transformation workflow in a principled way is thus of interest to many scientific disciplines, particularly areas such as medical imaging and brain mapping where ground truths are largely unknown or inaccessible. \textit{Differentiable programming} promises one potential resolution to this problem. A differentiable program can instantiate each block of a workflow as a neural network module that subsumes (a subset of) existing analytic options under reparameterisation. In other words, this differentiable program is a functional that interpolates over existing workflow configurations, relaxing the combinatorial problem of principled workflow design into one that can be optimised locally using gradient methods.

\begin{figure}
\includegraphics[width=\linewidth]{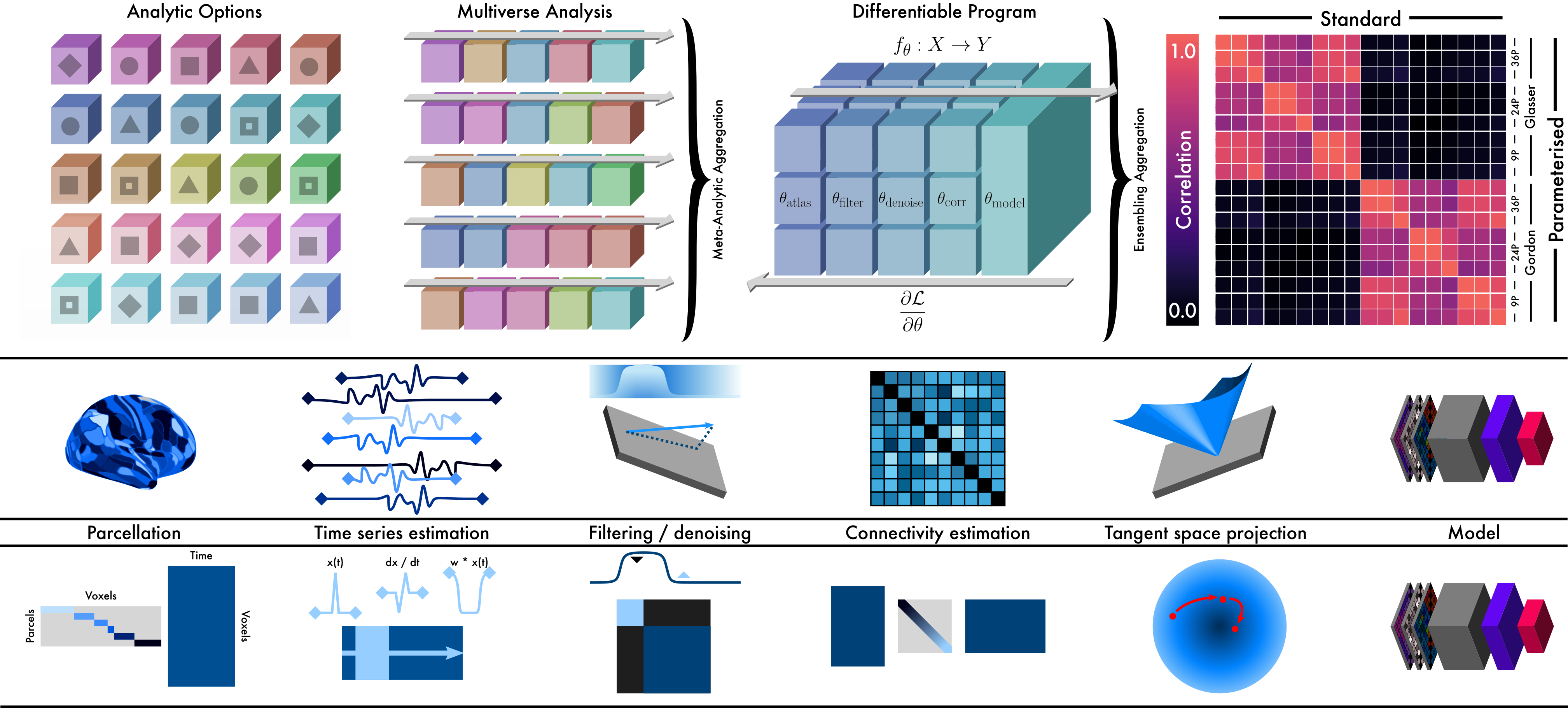}
\caption{Problem setting and aspirational overview. Details in text.}
\label{fig:overview}
\end{figure}

Functional connectomics is the enterprise of creating, refining, and explaining whole-brain maps of synchrony and statistical dependence in order to develop an understanding of how neuronal populations communicate with one another \citep{smith_functional_2013}. The standard functional connectivity workflow (Figure \ref{fig:overview}, \textit{Bottom}) begins with a preprocessed blood oxygenation level-dependent (BOLD) time series---a proxy for underlying neural activity \citep{logothetis_neurophysiological_2001}. A \textit{parcellation} block first reduces the dimension of this input, and the estimated parcel-wise time series are then \textit{denoised}. Next, some measure of \textit{connectivity} among the denoised time series is estimated; this connectivity matrix might be projected from the positive semidefinite cone\citep{golland_transport_2014} before it is passed to a final \textit{model} that is fit to some research objective. Differentiable programming enables gradients to propagate back from the model to reconfigure the parameters of the workflow itself.

Functional connectomics is an attractive test bed for prototyping differentiable programs because nearly all atoms of the workflow described above are either immediately differentiable or have differentiable relaxations. To demonstrate this, we implemented 18 different functional connectivity workflow configurations using different combinations of parcellations, denoising schemes, and filters. We then developed an open source PyTorch-based \citep{NEURIPS2019_9015} software library, \texttt{hypercoil}\footnote{\texttt{https://hypercoil.github.io/}}, to facilitate differentiable programming for functional connectomics. We used this library to parameterise neural network modules that replicated exactly each of the standard pipelines, as evidenced by perfect correspondence between the connectomes they output (Figure \ref{fig:overview}, \textit{Top right}). We also conducted a series of experiments, detailed below, as a proof of concept for the differentiable programming paradigm in functional connectomics. All experiments in the present manuscript were implemented using functionality freely available in \texttt{hypercoil}.

\section{Related work}

Parameterising the entire neuroimaging software stack as a neural network was first proposed by \citet{vilamala_towards_2016}, who designed a module that could learn an optimal spatial smoothing kernel for fMRI data. The current work was directly inspired by recent comprehensive combinatorial evaluation of functional connectivity workflows in the prediction setting \citep{pervaiz_optimising_2020,dadi_benchmarking_2019}. \citet{churchill_optimizing_2017} introduced an algorithm for adaptive neuroimaging workflow optimisation using a cross-validation framework. \citet{dafflon_neuroimaging_2020} developed a learning method for efficiently estimating an approximate map of the space of workflow configurations and of their performance on a prediction objective.

Differentiable programming is not the only promising approach for resolving the problem of principled workflow design. A complementary paradigm, \textit{multiverse analysis} \citep{Botvinik-Nezer2020-us}, is based on evidence that, although results gleaned from individual workflow configurations are often brittle and workflow-specific, meta-analytic aggregation across the ``multiverse'' of workflow configurations can separate workflow-sensitive results from those that are robust and reproducible \citep{Botvinik-Nezer2020-us}. Differentiable programming enables a complementary kind of aggregation: ensembling across a ``multi-barreled'' workflow whose parallel program blocks can learn complementary information in service of a research objective (Figure \ref{fig:overview}, \textit{Top}).

\section{Experiments}

\paragraph{Functional parcellation} We begin with a differentiable implementation of the first workflow block, brain parcellation. The delineation of functional subunits of the brain is among the longest standing problems in neuroscience. Practically, brain parcellation is also a critical dimension-reducing step that reduces the computational requirements of all downstream blocks. For BOLD fMRI data, the objective of the parcellation problem is to learn a mapping from the vertex-wise time series $\mathbf{T_i} \in \mathbb{R}^{v \times t}$ to the parcel-wise time series $\mathbf{T_o} \in \mathbb{R}^{p \times t}$ with minimal loss of information; in practice, this mapping is usually operationalised as either a linear parcellation matrix $\mathbf{A} \in \mathbb{R}^{p \times v}$ or a linear projection \citep{beckmann_group_2009,dadi_fine-grain_2020}. Here we take the former approach due to its simplicity, although our method is easily extended to the latter. Each entry $\mathbf{A}_{ij}$ in the parcellation matrix encodes the model's estimate of the probability that vertex $j$ is assigned to parcel $i$. We initialise each column of $\mathbf{A}$ by first sampling from a Dirichlet distribution and then log-transforming the Dirichlet samples. During each forward pass, the parcellation logits are projected to the probability simplex using a softmax mapping.
\begin{table}
\caption{The four loss terms of the differentiable temporal-spatial clustering (dTSC) objective. Details online at \texttt{https://hypercoil.github.io/loss.html}.}
\begin{center}
    \begin{tabular}{ r c c }
        \toprule
        & \thead{\textbf{Spatial}} & \thead{\textbf{Temporal}} \\
        \midrule
        \textbf{Within} & \makecell{\textit{Compactness} \\ $\mathbf{A} \circ \left\|\mathbf{C} - \frac{\mathbf{AC}}{\mathbf{A1}} \right\|_{\mathrm{cols}}$} & \makecell{\textit{Second Moment} \\ $\left[ \mathbf{A} \circ \left (\mathbf{T_i} - \frac{\mathbf{T_o}}{\mathbf{A 1}} \right )^2  \right] \frac{\mathbf{1}}{\mathbf{A 1}}$} \\ \\
        \textbf{Between} & \makecell{\textit{Dispersion} \\ $-\sum_{i, j} \mathrm{d}\left( \frac{\mathbf{A_i C}}{\mathbf{A_i 1}} - \frac{\mathbf{A_j C}}{\mathbf{A_j 1}} \right)$} & \makecell{\textit{Determinant} \\ $-\log \det \mathrm{corr} ( \mathbf{T_o} )$} \\
        \bottomrule
    \end{tabular}
\end{center}
\label{table:dtsc}
\vspace{-2mm}
\end{table}

\begin{figure}
\includegraphics[width=\linewidth]{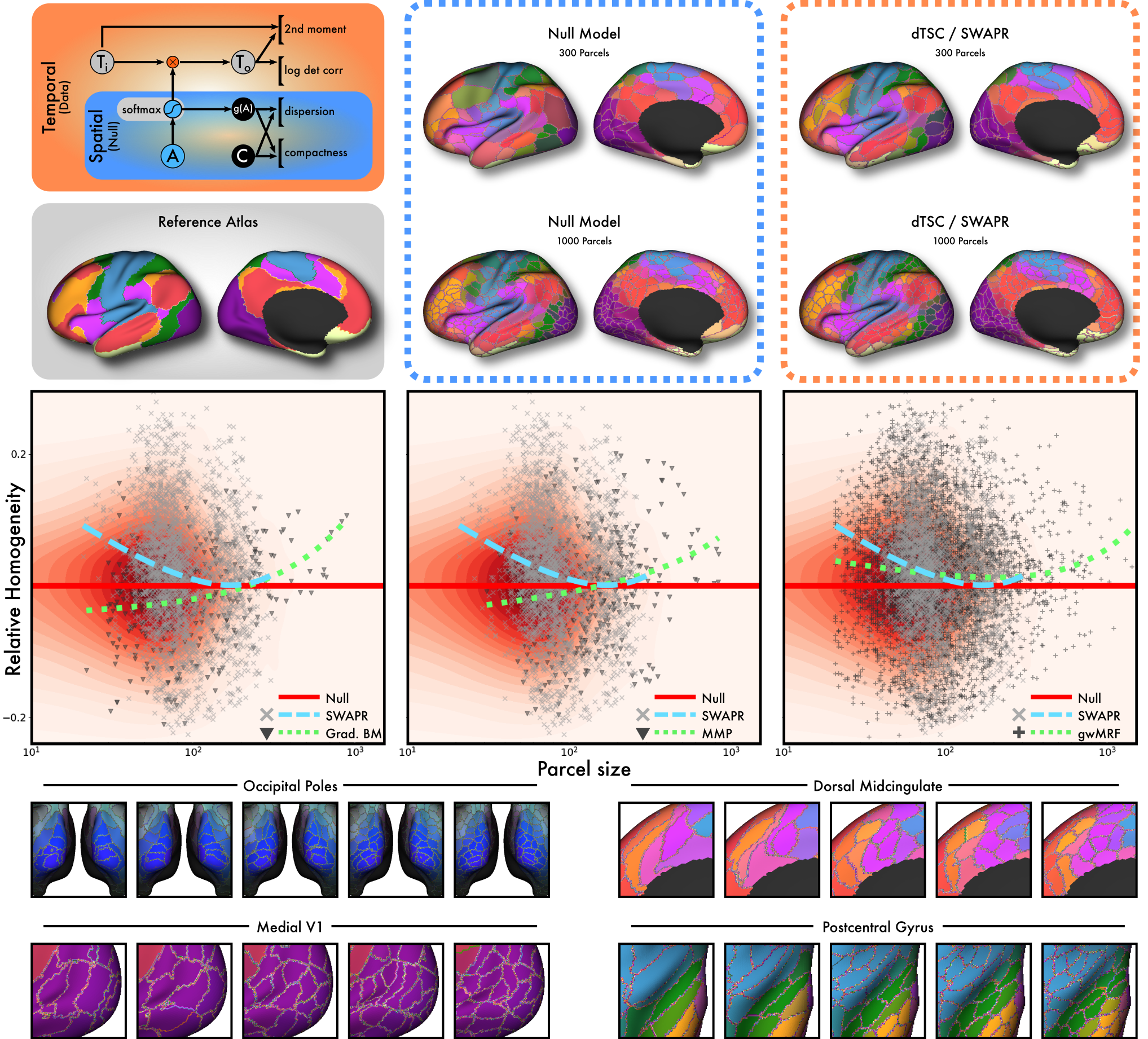}
\caption{dTSC / SWAPR parcellation. \textit{(Top left)} Schematic of the dTSC model. \textit{(Top)} The model recovers reference functional boundaries \citep{yeo_organization_2011} with success relative to a null model. \textit{(Middle)} Quantitative assessment of parcellation homogeneity (higher is better). \textit{(Bottom)} Consistency and difference across five parcellation resolutions in four representative regions.}
\label{fig:swapr}
\end{figure}

To learn the parcellation matrix $\mathbf{A}$, we combine four terms into a differentiable temporal-spatial clustering (dTSC) objective (Figure \ref{fig:swapr}, \textit{Top left}; Table \ref{table:dtsc}). A \textit{compactness} loss penalises the distance from each vertex in a parcel to the parcel's centre of mass using vertex coordinates $\mathbf{C}$, thereby promoting spatially compact parcels. A complementary \textit{dispersion} objective promotes spatial separation of different parcels' centres of mass. All distances are computed as spherical geodesics after projecting data from each cortical hemisphere onto a spherical mesh \citep{fischl_cortical_1999,robinson_msm_2014}.

To minimise information loss and learn functionally uniform parcels, we wish to assign vertices with similar signals to the same parcel and vertices with different signals to different parcels. Accordingly, in addition to the two spatial terms, temporal loss terms promote parcel homogeneity. A \textit{second moment} objective quadratically penalises parcels for including vertices whose time series differ from the parcel's time series. Finally, we promote temporally independent parcels by placing a penalty on the negative \textit{log-determinant} of the correlation matrix among parcel time series. Our approach is readily modified to produce null parcellations: ablating the temporal loss terms leaves a model that learns entirely from spatial relationships, without using any information from the time series data.

\begin{figure}
    \centering
    \subfloat[Parcel assignment uncertainty is modulated by a progressive penalty on the distribution entropy. The paired vertical lines denote training cycles during which the entropy multiplier is cascaded upward.]{
        \includegraphics[width=.48\hsize]{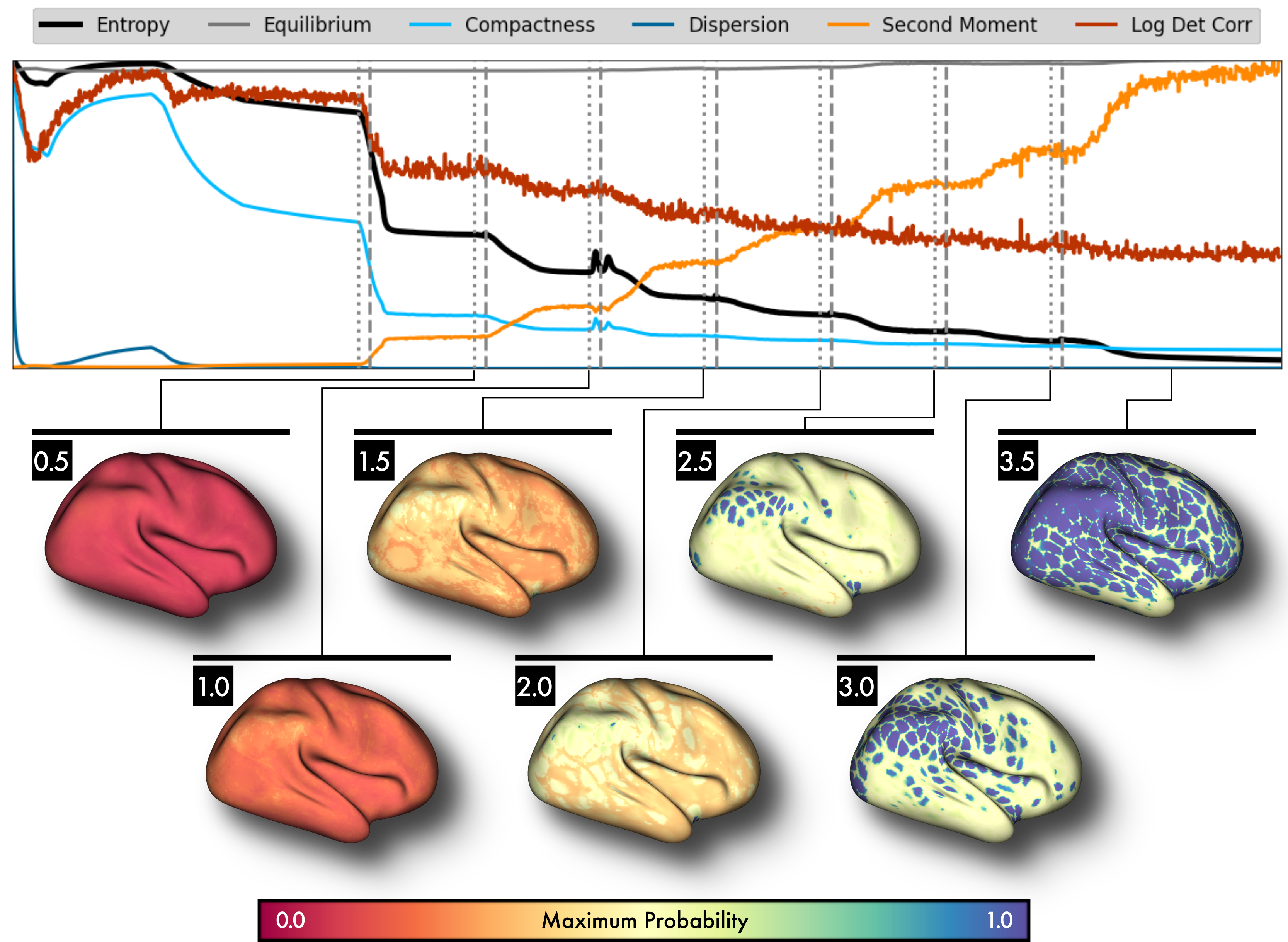}
    }\qquad
    \subfloat[The learned parcellation proceeds from distributed functional modes to deterministic areal parcels. Here, we show four exemplar parcels from the right cortical hemisphere of a 666-region parcellation at the end of each of four stages of the entropy cascade.]{
        \includegraphics[width=.43\hsize]{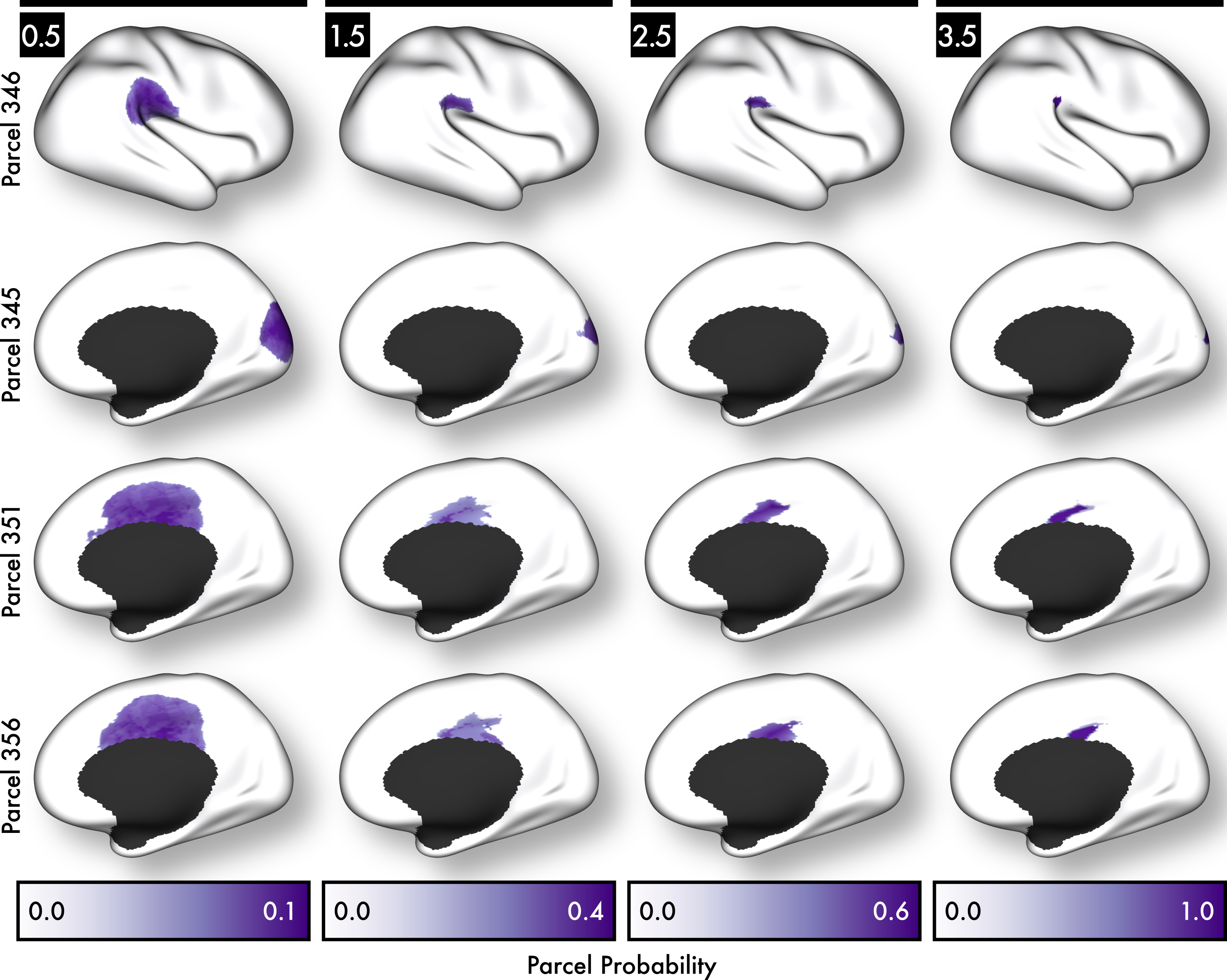}
    }
    \caption{Learning a parcellation using the dTSC / SWAPR model with a cascading entropy loss.}
    \label{fig:cascade}
\end{figure}

We supplement the dTSC objective with three simple regularisations. First, to obtain parcels of approximately equal size, we impose an L2 parcel equilibrium penalty. Second, we enforce approximate symmetry across cortical hemispheres by tethering each parcel's centre to an analogue in the opposite hemisphere using another L2 penalty. Third, to obtain deterministic parcels, we penalise the entropy of each vertex's parcel assignment distribution. Because a strong entropy penalty can lead the model to fixate irreversibly on its maximum assignment, we begin with a small multiplier for the entropy term and progressively cascade it upward over the course of training. The cascading entropy loss induces parcel probabilities to converge toward a deterministic assignment, at the cost of worsening parcel homogeneity as reflected by a growing second-moment term (Figure \ref{fig:cascade}a, \textit{Top}), with boundary regions slowest to converge (Figure \ref{fig:cascade}a, \textit{Bottom}, entropy multipliers denoted in black boxes).

Because the boundaries of brain subsystems vary substantially between individuals \citep{laumann_functional_2015,gordon_precision_2017}, we use weight averaging to smooth the objective for our parcellation. In particular, we modify the stochastic weight averaging (SWA; \cite{izmailov_averaging_2019}) algorithm for better compatibility with entropy penalties. Our modification, SWA with Parameter Revolution (SWAPR), yields parcels that qualitatively better match the boundaries of reference group-averaged brain subsystems (\cite{yeo_organization_2011}; Figure \ref{fig:swapr}, \textit{Top}).

We used the dTSC objective with SWAPR to train 5 different brain parcellation models corresponding to 5 spatial scales (300, 400, 666, 800, 1000 cortical parcels) using 2457 BOLD images from the Human Connectome Project (HCP) dataset \citep{van_essen_wu-minn_2013}. In comparison with a spatial null model, the dTSC / SWAPR parcellations better reflected established contours of the brain's canonical functional subsystems (Figure \ref{fig:swapr}, \textit{Top}) across all spatial resolutions, with differences particularly evident at coarser resolutions (e.g., 300 parcels). On average, the deterministic parcels learned by our model also achieved better signal homogeneity than comparably sized parcels from the null model and performed competitively with state-of-the-art methods (gwMRF: gradient-weighted Markov Random Field \citep{schaefer_local-global_2018}; MMP: multimodal parcellation \citep{glasser_multi-modal_2016}; Grad. BM: gradient boundary mapping \citep{gordon_generation_2016}) in held-out data (Figure \ref{fig:swapr}, \textit{Middle}).

Across spatial resolutions, we observed consistency and difference in parcel structure (Figure \ref{fig:swapr}, \textit{Bottom}). Although some areal boundaries (notably of the occipital visual cortex and medial V1) were reproduced with high fidelity across resolutions, we qualitatively found greater consistency in the orientation of local axes of signal similarity and difference, as reflected in the eccentricities and orientations of parcel contours. This pattern suggests a limitation of areal parcellation schemes such as the one we introduce; parcellations based on connectopic gradients (e.g., \cite{tian_topographic_2020}) offer a potential direction for future improvement.

Another interesting insight can be derived from the cascading entropy penalty used to train the parcellation: at each stage of the cascade (labelled with the entropy multiplier in Figure \ref{fig:cascade}b), the spatial extent of parcels decreases, and their maximum assignment probability increases toward unity. Qualitatively, this is reflected in a shift from spatially overlapping, probabilistic modes of brain activity (e.g., \cite{dadi_fine-grain_2020}) toward deterministic areal assignments. In particular, the shown areal parcels 351 and 356 begin with near complete overlap and substantially differentiate only in the final stages of the cascade. By evaluating parcel differentiation over the course of the cascade, the progression of training can potentially be used to investigate hierarchically nested modes of brain function \citep{pines_dissociable_2022}. As complementary information might be offered by overlapping probabilistic modes and circumscribed parcels, the stopping entropy also becomes a hyperparameter of the differentiable program, which can be ensembled over.

Our approach differs from most previous work in that it is designed to fit entirely in GPU memory and (with some refinement) integrate into a unified differentiable workflow. Principally, we accomplish this (i) by using only global information about the absolute spatial position of each vertex, while many competing methods maintain full adjacency graphs of neighbours, often performing expensive label propagation updates at every step, and (ii) by using only losses that operate directly on the $v \times t$ vertex-wise BOLD time series, bypassing the need to compute the complete $v \times v$ connectivity matrix, which is computationally slow and prohibitively expensive to store all at once in GPU memory. Our deterministic parcels are accordingly not as circumscribed as those found by methods that do incorporate local information (e.g., \cite{gordon_generation_2016,glasser_multi-modal_2016,schaefer_local-global_2018}), although this is easily remedied using (nondifferentiable) automatic postprocessing options such as connected component identification. However, we find that (even subject to these constraints) we can achieve parcel homogeneities that are competitive with SOTA in held-out data (Figure \ref{fig:swapr}, middle row).

\paragraph{Artefact removal}
Next, we implement a differentiable module for denoising BOLD time series. The movement of subjects during a scan introduces artefactual fluctuations into the BOLD signal \citep{power_spurious_2012}, which typically inflate estimates of functional connectivity. To make matters worse, motion artefact is correlated with measures of scientific interest, such as subject age \citep{satterthwaite_impact_2012}. In the functional connectivity workflow, denoising is typically implemented by residualising the BOLD time series with respect to a \textit{confound model} that is thought to approximately explain structured sources of artefact. We adopt this framework here.

\begin{figure}
    \centering
    \caption{A shallow denoising model (RFNN) trained on the QC-FC loss outperforms SOTA methods in held-out data.}
    \begin{minipage}[t]{.42\linewidth}
    \subfloat[The RFNN model learned to remove a single nuisance regressor from BOLD images so as to reduce motion-related variance.]
        {\includegraphics[width=\linewidth]{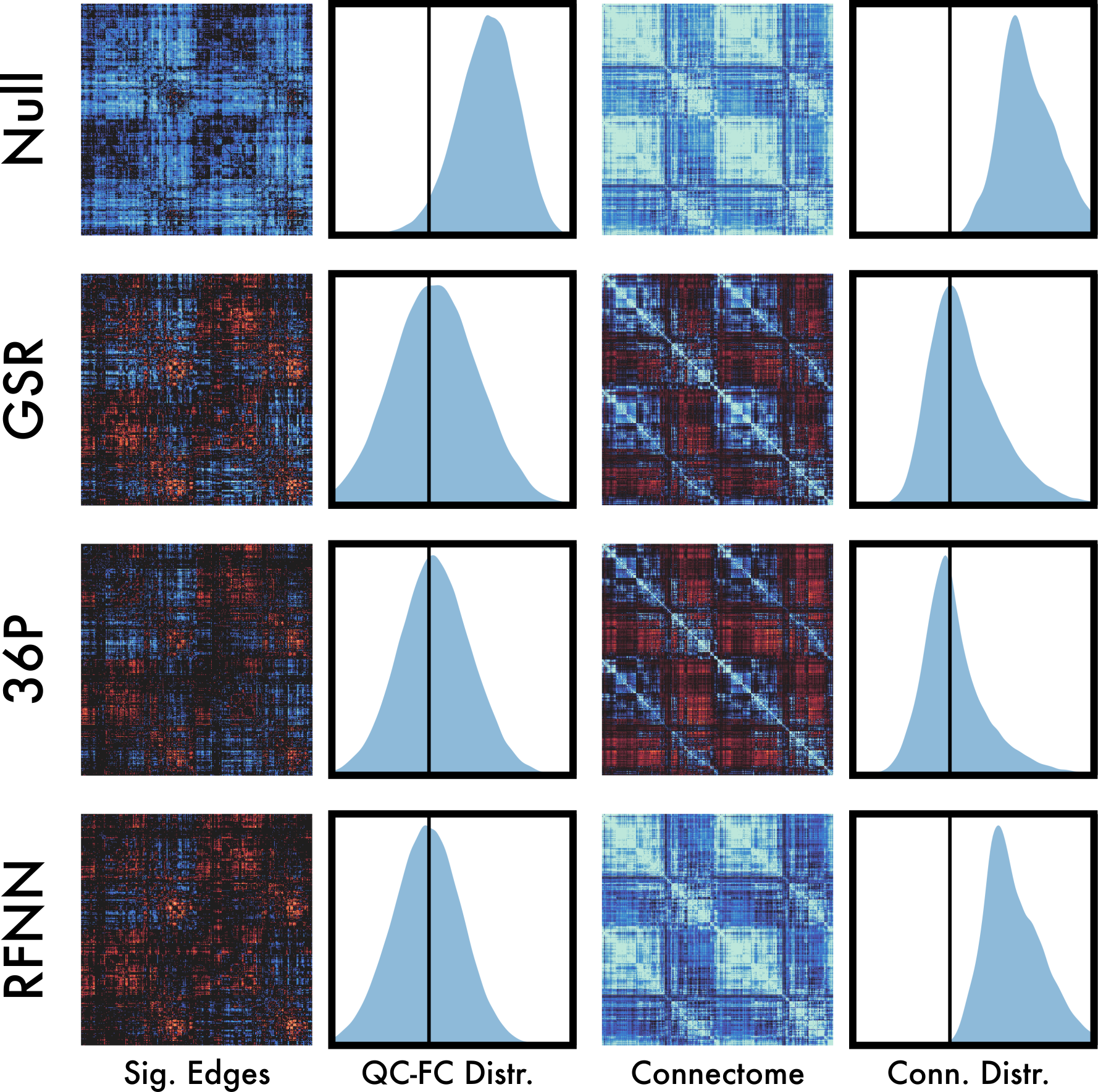}}
    \end{minipage}
    \hfill
    \begin{minipage}[t]{.5\linewidth}
    \subfloat[QC-FC benchmark results (lower is better).]
    {
        \begin{tabular}{l r r}
            \toprule
            Model & {Abs. Med. Corr.} & N. Sig. Edges \\
            \midrule
            Null & 0.16640 & 50765 \\
            GSR & 0.09036 & 14558 \\
            36P & 0.07487 & 11099 \\
            RFNN & \textbf{0.07185} & \textbf{6439} \\
            \bottomrule
        \end{tabular}
    }\\
    \subfloat[Top \textit{a priori} confounds that loaded onto the RFNN model were global signal (GS)-related.]
        {\includegraphics[width=\linewidth]{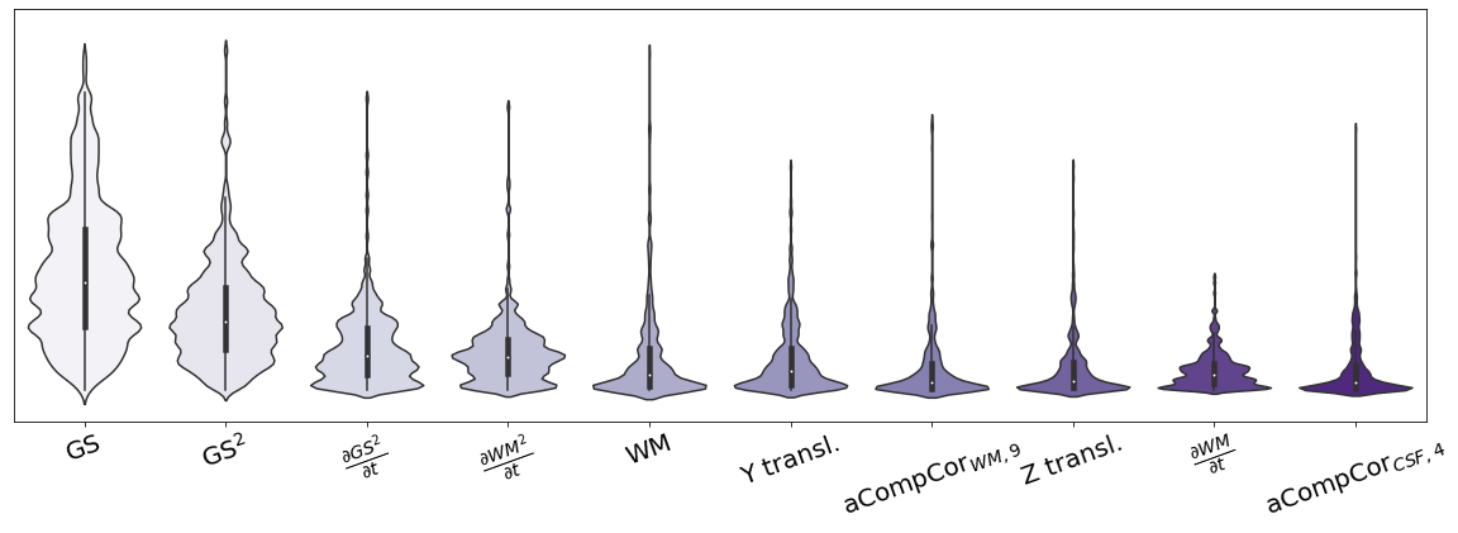}}
    \end{minipage}
    \label{fig:denoise}
\end{figure}

In particular, let $\Sigma_{XX}$ be the covariance matrix of BOLD time series, $\Sigma_{YY}$ be the covariance matrix of confound time series, and $\Sigma_{XY}$ be the matrix of covariances between BOLD and confound time series. We operationalise the residual functional connectivity as the \textit{conditional covariance}
\begin{align*}
    \Sigma_{X|Y} &= \Sigma_{XX} - \Sigma_{XY} \Sigma_{YY}^{-1} \Sigma_{XY}^\intercal
\end{align*}
This is equivalent to the Schur complement of the confound covariance, and also to the covariance of BOLD data that have been residualised with respect to a linear least-squares fit of the confounds. \textit{Ceteris paribus}, a more parsimonious confound model---one with fewer time series---is thought to be better because it removes fewer degrees of freedom from the BOLD data \citep{pruim_evaluation_2015}. Thus, in this experiment, our objective is to remove as much motion-related variance as possible from the BOLD data using only a single learned confound time series.

We pursue this objective by parameterising a simple, shallow neural network model (RFNN) to learn a single linear combination of 57 confound time series selected \textit{a priori} for their demonstrated efficacy in denoising \citep{ciric_benchmarking_2017}. These include direct estimates of subject movement; mean signals from high-noise tissue compartments (white matter---WM, cerebrospinal fluid---CSF); the overall global signal across the entire brain (GS); and localised signals obtained using singular value decompositions of WM and CSF (CompCor; \cite{behzadi_component_2007}). We parameterise the RFNN to also learn 5 response functions; the RFNN is permitted to select any linear combination of the 57 \textit{a priori} confound time series and their $285 = 5 \times 57$ response function convolutions.

To train the RFNN to optimally remove motion artefact, we introduce the \textit{QC-FC} loss function, a differentiable implementation of the standard QC-FC benchmark for residual motion artefact \citep{power_methods_2014}. The QC-FC loss computes the correlation between a gross estimate of subject motion (QC) and each edge of the functional connectivity matrix (FC) across the batch dimension. Thus, QC-FC loss goes to zero when subject movement and functional connectivity are uncorrelated.

In a held-out sample of 351 BOLD images from the HCP dataset (Figure \ref{fig:denoise}), the RFNN model outperformed both the top-performing \textit{a priori} one-confound model (GSR, global signal regression) and a 36-confound model (36P, a superset of GSR) that has previously been demonstrated to give SOTA performance \citep{ciric_benchmarking_2017}. Both the number of connections significantly related to motion (Sig. Edges; $p < 0.01$, uncorrected) and the median of the size of motion effects (QC-FC Distr.) were reduced relative to all evaluated \textit{a priori} models. We caution that this superior performance reflects a limitation of the benchmarks: the functional connectome denoised using the RFNN did not feature the zero-centred correlations that are a hallmark of successful GSR-based denoising \citep{murphy_impact_2009}. It is possible that this reflects the specificity of the learned model for removing motion-related variance---by contrast, GSR-based models also effectively remove physiological artefacts related to respiratory and cardiac processes, which typically inflate global estimates of connectivity \citep{power_sources_2017}. With this and additional benefits of GSR in consideration \citep{li_global_2019}, we strongly recommend continued use of GSR in functional connectivity workflows; future work will address possible integration of RFNN-like models with GSR. Notably, when we computed the extent to which \textit{a priori} confounds captured variance in the RFNN confound, the top three loaded \textit{a priori} confounds were all GSR-related: the global signal (GS), the square of the global signal ($\mathrm{GS}^2$), and the temporal derivative of the square ($\frac{\partial \mathrm{GS}^2}{\partial t}$) (Figure \ref{fig:denoise} \textit{Top}, distribution across images).

\begin{figure}
    \centering
    \includegraphics[width=\linewidth]{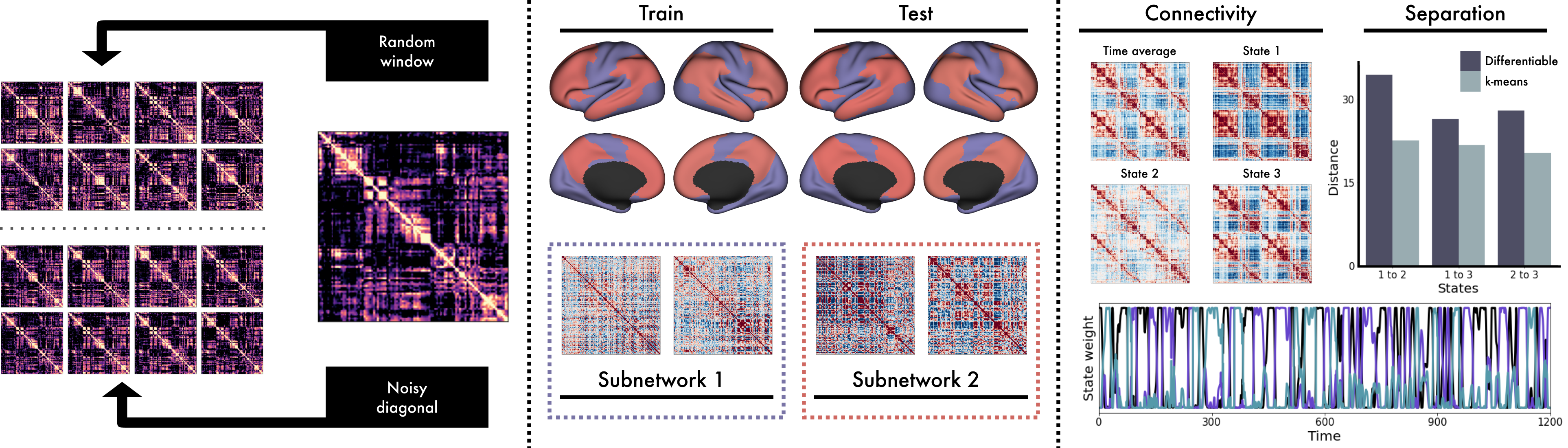}
    \caption{Results of covariance experiments. \textit{Left}, Methods for augmenting covariance data. \textit{Centre}, Clustering on the time-by-time covariance matrix partitions the brain into unimodal and higher-order subnetworks with distinct temporal reconfiguration profiles. \textit{Right}, Clustering on the parcel-by parcel covariance matrix partitions each time series into connectivity states.}
    \label{fig:cov}
\end{figure}

\paragraph{Covariance clustering}
We finally turn our attention to covariance estimation. The functional connectome is typically estimated using a derivative of covariance, such as Pearson correlation, among all pairs of parcel time series. The usual definition of empirical covariance among a set of time series $\mathbf{T}$ can be generalised via parameterisation by the (typically diagonal) weight matrix $\mathbf{\Theta}$: $\mathbf{C_\Theta} = \frac{1}{n - 1} \left(\mathbf{T} - \mathbf{\bar{T}}\right) \mathbf{\Theta} \left(\mathbf{T} - \mathbf{\bar{T}}\right)^\intercal$. This parameterised form finds an application in data augmentation: setting the weights $\theta_{ii}$ along the diagonal to random nonnegative integers whose sum is the total number of observations is equivalent to a resample. Relaxing this constraint to require only a nonnegative support and a mean of 1, satisfied \textit{inter alia} by noise sampled from an appropriately chosen gamma or truncated normal distribution, leads to a simple method for augmenting covariance datasets that complements random windowing of the input time series (Figure \ref{fig:cov}, \textit{Left}).

In this proof of concept, however, our focus is on the problem setting of clustering. Specifically, how can we cluster observations such that the covariance matrices estimated from different clusters are, for some measure of separation, maximally different? This nonconvex maximisation problem has two immediate applications in functional connectomics: \textit{subnetwork detection} and \textit{state detection}. To further elaborate, given the time series matrix $\mathbf{T} \in \mathbb{R}^{p \times t}$, we can obtain two covariance matrices, the $p \times p$ connectome matrix of inter-parcel correlations and the $t \times t$ matrix of correlations among brain activity profiles across time (e.g., \cite{medaglia_brain_2018}). For the $p \times p$ connectome, observations correspond to time points, and clustering them to produce maximally distinct connectomes can be interpreted as brain state detection. For the $t \times t$ matrix, observations correspond to different brain parcels; clustering them is a form of subnetwork detection.

For each problem setting, we train a clustering model to learn $\Theta \in \mathbb{R}^{c \times p}$ or $\mathbb{R}^{c \times t}$, where $c$ is the number of clusters, here selected \textit{a priori}. We instantiate a clustering loss to maximise the total L2 dispersion among the covariance matrices corresponding to different detected clusters. As in the parcellation problem, we penalise the entropy to promote deterministic assignment of each parcel or time point to a single subnetwork or state. For the state detection problem, we also encourage the model to learn persistent states by imposing an L2 smoothness penalty on the backward difference of $\mathbf{\Theta}$. Because state detection (and numerous other neuroimaging applications, such as subject-specific parcellation) requires learning a unique time course for each data instance, the \texttt{hypercoil} package also includes extensions of PyTorch optimisers that accept ephemeral, instance-specific parameter groups for optimisation and optionally clear them from memory after they have been updated.

The results of the clustering experiment in subsamples of the HCP dataset are shown in Figure \ref{fig:cov}. Clustering $t \times t$ covariances into $c = 2$ subnetworks divides the brain along a unimodal (blue) to higher-order (red) axis (Figure \ref{fig:cov}, \textit{Centre}); this clustering is replicable across data splits. We show the $t \times t$ covariance matrices for 2 example subjects, illustrating the distinct dynamic profiles of the two subnetworks. Clustering $p \times p$ covariances into $c = 3$ states yields three whole-brain connectivity states that are each distinct both from one another and from the time-averaged connectome (Figure \ref{fig:cov}, \textit{Right}). The learned assignment of time frames to the three states is plotted for an example subject. In contrast with the most commonly used state detection methods (e.g. \cite{allen_tracking_2014}), the method we apply does not require estimating the covariance over a sliding window; it instead directly learns to assign time frames to states. Compared with a classical approach that uses $k$-means clustering \citep{allen_tracking_2014}, our differentiable method learns superior separation of each pair of detected brain states in our test case.

\paragraph{Community detection}
As a second proof of concept for connectome estimation, we combine the parameterised covariance function described in the previous section with community detection. The objective of community detection is to learn a partition of graph vertices (here, brain parcels) into \textit{communities} whose members preferentially connect to one another (e.g., \cite{blondel_fast_2008}). In connectomics, this corresponds to identification of modular subsystems of brain function. By combining community detection with the parameterised covariance described previously, we can learn dynamic time courses that track the modularisation and demodularisation of these functional subsystems.

Early algorithms for community detection were neither convex nor differentiable, although more recently neural networks have been developed to learn community partitions \citep{su_comprehensive_2022}. Here, we opt for a more direct approach, because functional connectome matrices typically have fewer vertices than the large-scale graphs for which deep methods were developed. Our approach also differs from most algorithms for dynamic community detection (e.g., \cite{thompson_identification_2019,martinet_robust_2020}) in that existing algorithms typically aim to characterise the temporal evolution of community boundaries, whereas the objective of our approach is to delineate epochs during which communities coalesce and dissipate.

We train our model using a differentiable relaxation of the Girvan-Newman modularity objective \citep{newman_finding_2004}, which indexes the ability of a proposed community structure to explain the arrangement of edges in a real graph relative to a null model. Formally, let $\mathbf{A} \in \mathbb{R}^{v \times v}$ be the graph adjacency matrix and $\mathbf{C} \in \mathbb{R}^{v \times c}$ be a proposed assignment of $v$ vertices to $c$ communities. The \textit{modularity matrix} $\mathbf{B} = \mathbf{A} - \gamma \mathbf{P}$ expresses the extent to which edges in the observed graph deviate from expectation under the null model $\mathbf{P}$, subject to a resolution hyperparameter $\gamma$. Here, we use $\gamma = 5$ and the original Girvan-Newman null model, $\mathbf{P}_{GN} = \frac{\mathbf{A11}^\intercal \mathbf{A}}{\mathbf{1}^\intercal \mathbf{A} \mathbf{1}}$, which can be interpreted as the expected weight of connections between each pair of vertices if all existing edges are cut and then randomly rewired. The modularity objective is then $\mathcal{L}_Q = \mathbf{1}^\intercal (\mathbf{H} \circ \mathbf{B}) \mathbf{1}$, where $\circ$ denotes the Hadamard product and $\mathbf{H} = \mathbf{C C}^\intercal$ is the community coaffiliation matrix. Our goal is to learn a $\mathbf{C}$ that maximises the modularity objective. If we constrain all entries in $\mathbf{C}$ to $\{0, 1\}$, this relaxation converges to the original definition of modularity \citep{newman_finding_2004}. To maintain differentiability, we instead learn the logits of $\mathbf{C}$, which we pass through a softmax, permitting community assignments to vary continuously in the probability simplex.

\begin{figure}
    \centering
    \includegraphics[width=0.7\textwidth]{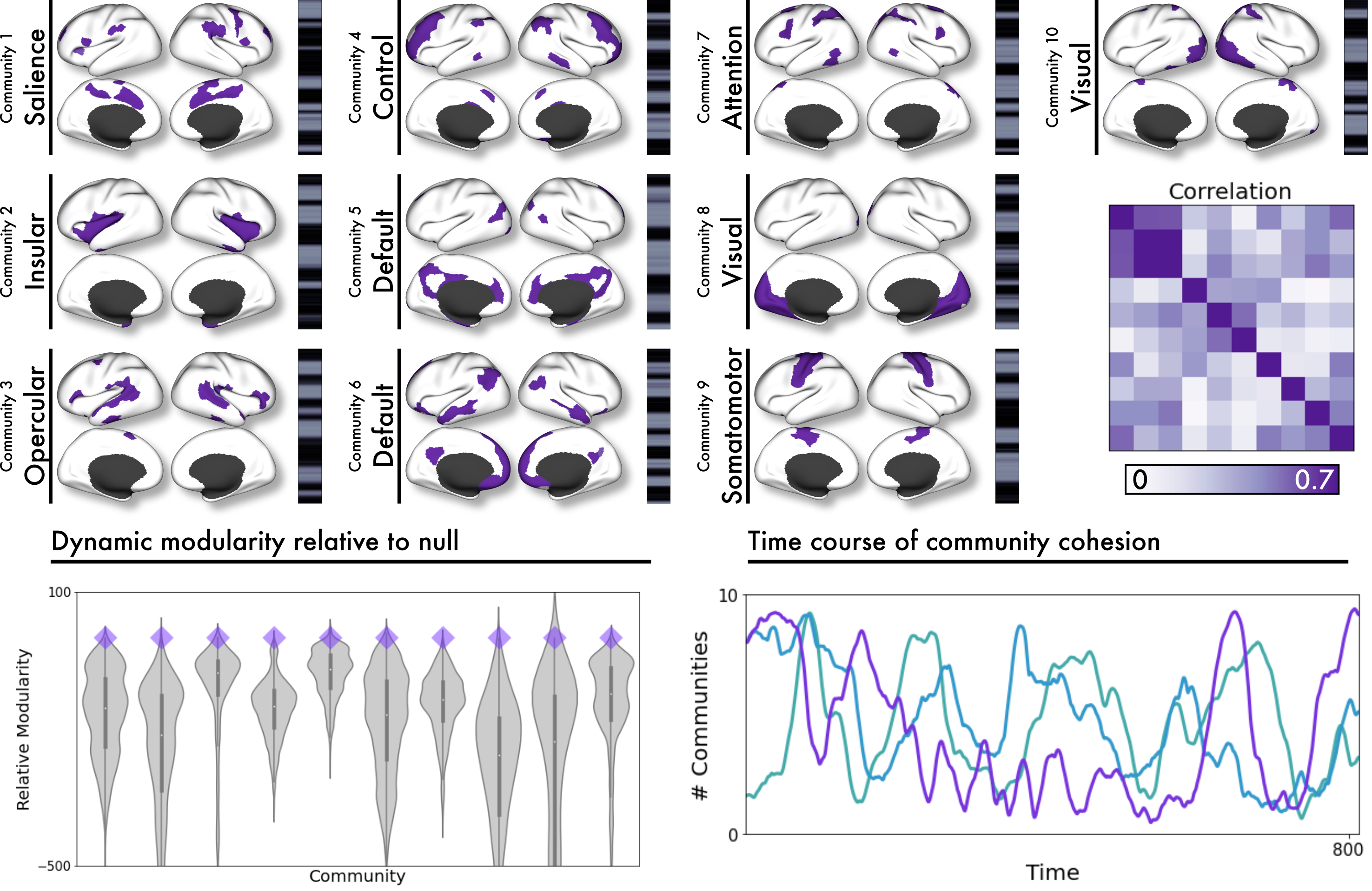}
    \caption{A community detection model learns to track the modularisation and demodularisation of brain subsystems.}
    \label{fig:community}
\end{figure}

The modularity objective is applied to the time-averaged connectome and combined with time-selective modularity objectives applied separately for each community. Specifically, for each community $\mathcal{C}_i$, we let $\mathbf{A}_{\mathcal{C}_i}$ be the covariance parameterised by a learnable time course $\mathbf{\Theta}_{\mathcal{C}_i}$ of the community's modularisation, and $\mathbf{H}_{\mathcal{C}_i} = \mathbf{C}_i \mathbf{C}_i^\intercal$. As before, we regularise $\mathbf{\Theta}_{\mathcal{C}_i}$ using a smoothness/persistence loss.

Training this unsupervised model on 30 parcellated time series from the Midnight Scan Club (MSC) dataset \citep{gordon_precision_2017}, we identify a set of dynamic communities that accord well with previously characterised brain subnetworks. At the top of Figure \ref{fig:community}, we show the learned spatial distributions $\mathbf{C}_i$ of each community and an example time course $\mathbf{\Theta}_{\mathcal{C}_i}$ of the community's modularisation (grey: present, black: absent) in a single BOLD image. Relative to a null model obtained by randomly shifting modularisation time courses, the learned model consistently obtained better modularity (Figure \ref{fig:community}, \textit{Bottom left}). We also find that communities tend to coalesce and dissipate in tandem (correlation matrix and time course at the right of Figure \ref{fig:community}), consistent with previous work that characterised transient ``excursions'' of connectivity using a sliding-window approach \citep{betzel_dynamic_2016}.

\section{Discussion}

Our experimental results demonstrate the promises of a fully differentiable software stack in the domain of functional connectomics. All experiments were implemented using the PyTorch-based \texttt{hypercoil} software library that we introduce and freely release. This manuscript additionally functions as an open invitation to members of the community with an interest in applying differentiable programming to brain mapping to contribute to further research and development of \texttt{hypercoil}.

We must remark that a substantial prior body of work has established limited utility for deep neural networks in the setting of brain mapping (e.g., \cite{pervaiz_optimising_2020,he_deep_2020,schulz_different_2020,thomas_classifying_2020}, but see also \cite{abrol_deep_2021}). Importantly, this work has largely focused on one mode of scientific understanding---prediction. We believe, based on the proof of concept presented here, that the greatest promise of the system we introduce lies not in prediction but in developing other modes of understanding about neuroscience and the design of scientific workflows.

There additionally exists a central philosophical objection against the differentiable programming paradigm. The argument goes as follows: differentiable programming does not in fact resolve the problem of workflow design; instead, it is merely a re-delegation of the workflow design burden from selection of analytic options to specification of workflow hyperparameters. Different hyperparameter configurations (e.g., the balances of loss and regularisation multipliers) are inherently going to produce different ``optimal'' workflows. Although it can be argued that the hyperparameter selection process sometimes offers researchers more transparent control over workflow design, because a numerical objective is more intuitively related to a research objective, this is often not the case. Although responses to this argument (e.g., mapping and understanding the most relevant subsets of the hyperparameter space; strides in automating the process of hyperparameter optimisation, e.g. \cite{he_automl_2021}) are under active development in the machine learning world, this does not fully address doubts. Nevertheless, even if a differentiable program does not completely address the problem of principled workflow optimisation, it has the potential to provide both an engine for new discoveries and an informative complement to multiverse analysis.

\printbibliography

\newpage
\hrule height 4pt
\vspace{0.09in}
\begin{center}
    {\LARGE\bf Differentiable programming for functional connectomics\\ \large{Supplemental matter}\par}
\end{center}
\vspace{0.09in}
\hrule height 1pt
\vspace{0.09in}

\renewcommand{\thefigure}{S\arabic{figure}}
\renewcommand{\thetable}{S\arabic{table}}
\setcounter{figure}{0}
\setcounter{table}{0}

\begin{refsection}

\appendix
\section{Parcellation}

\paragraph{Dataset, preprocessing, and computational resources}
The openly available Human Connectome Project (HCP) dataset comprises scans from 1200 participants performing a number of in-scanner tasks. For each participant, the study acquisition protocols included 4 separate resting-state (task-free) scans, each acquired over the course of approximately 15 minutes of scanning using an accelerated multiband sequence with a sampling rate of 0.72 seconds \citep{van_essen_human_2012,van_essen_wu-minn_2013}. In reality, the protocols were incomplete for many participants.

To obtain our training and evaluation datasets, we first performed a random 12-way split on 1098 HCP subjects with resting-state fMRI data. We then selected the first 7 of these splits, comprising 2457 resting-state BOLD time series, to train our model. To create an evaluation set, we selected another split, from which a single 69-time series shard was randomly selected for homogeneity evaluation.

Data were preprocessed according to the minimal preprocessing pipeline of the Human Connectome Project \citep{glasser_minimal_2013}, and were acquired following a standardised protocol in accordance with the host institution's IRB \citep{van_essen_human_2012}. To better respect the topology of the cortex and to reduce the effects of inter-subject variability, we used time series projected onto a spherical surface \citep{fischl_cortical_1999} and aligned using the MSMAll algorithm \citep{robinson_msm_2014}. Before each time series was passed to the model, we performed several additional preprocessing steps. First, we projected each time series to the orthogonal complement of a subspace defined as the span of (i) a quadratic polynomial (to mitigate scanner drift artefact) and (ii) the average global signal across all voxels (following previous parcellation work, and in accordance with the overall demonstrated benefits of this approach; \cite{schaefer_local-global_2018,li_global_2019}). Next, we normalised each time series to a mean of 0 and a variance of 1. Finally, to reduce the memory footprint, we selected a random 500-sample window from each time series as input to the model.

The model was trained on a computer running Ubuntu Linux, using only a single commercial-grade (RTX 2080 Ti) GPU with 11 GB of RAM. Memory usage was reduced through serialisation of expensive loss computations (details in \textit{Loss} section below). Each of the 5 parcellations required approximately 18-24 hours of GPU time to train. Data were stored as \texttt{tar} shards for compatibility with the \texttt{webdataset} format \citep{aizman_high_2019}. Due to the large size of each BOLD time series, the most significant bottleneck during training was reading from the disk. Unfortunately, we found that data loaders were not well equipped to expedite this process for the large samples used, although it is likely that this was due to insufficient optimisation in sharding and worker deployment. We used a fixed batch size of 3 time series for training. To reduce disk I/O operations, we trained the model for 5 steps with each batch before sampling the next batch. The model was trained for a total 6000 steps using the Adam optimiser \citep{kingma_adam_2017}.

\paragraph{Loss function}
As detailed in the main text, the loss function we use comprises four clustering terms (spatial compactness of parcels, spatial dispersion among parcels, second moment of parcel time series, and negative log determinant of the correlation among parcel time series), as well as three regularisation terms (parcel equilibrium, vertex-wise distribution entropy, and inter-hemispheric spatial tether). We discuss here several details of the loss function implementation.

The broadcasting operations involved in the second moment term require computation of a large tensor for each hemisphere's BOLD time series, of dimension $v \times p \times t$, where $v$ is the number of vertices (order of 30000 per hemisphere), $p$ is the number of parcels (150-500 per hemisphere), and $t$ is the number of time frames (500 after random windowing). A tensor of this dimension was not able to fit into our GPU memory, so we performed a data flow serialisation to reduce the memory footprint of the second moment loss function. Specifically, we sliced the computation along the time axis, processing 3-10 time points per slice, looping over slices, and renormalising the slice-wise loss to adjust for slice size. The computational graph was immediately destroyed for each slice after computing the gradient. This enabled us to perform the necessary computation, and we verified equality of the accumulated gradients. In principle, this slicing approach should also enable us to work with the full $v \times v$ connectivity matrix (order of $60000 \times 60000$ for cortical-only or $90000 \times 90000$ for all coordinate time series) in loss computations used in future iterations of this method.

Through experimentation, we also made one further change to the second moment loss term: we removed the normalisation $\frac{\mathbf{1}}{\mathbf{A 1}}$, the use of which we found could yield extremely several large parcels that did not correspond with any well-known functional systems of the brain. With this change, the second moment term we use essentially reduces to a weighted mean squared error term, where each ``target'' is a parcel time series, each ``prediction'' is a vertex time series, and the weight assigned to each target-prediction pair is the assignment probability of the corresponding vertex to the corresponding parcel.

The negative log determinant term has the potential to go to infinity, for instance if parcel-wise time series are not independent. While the purpose of this term is to promote parcel independence, an exploding loss can irreversibly cause parameter values to go out of bounds when it is propagated back. To minimise the risk of this occurrence while preserving the efficacy of the determinant term, we added a small reconditioning term to the correlation matrix. For each entry along the diagonal of the correlation matrix, we randomly sampled i.i.d. noise from $\mathrm{Uniform}(0, 0.001)$ and added this noise to the diagonal. Although this reconditioning added further stochasticity to the determinant term, we observed an overall downward trend of the determinant loss during training.

Details of loss function definitions are available on the documentation hub at \texttt{https://hypercoil.github.io/loss.html}.

\paragraph{SWAPR algorithm}
SWAPR is a simple modification of Stochastic Weight Averaging (SWA; \cite{izmailov_averaging_2019}) that adds revolution of parameters between the averaged model and the data-facing model. The rationale for this modification follows: when entropy is sufficiently penalised, as it is in the final stages of the entropy cascade that we use, a data-facing model will eventually converge to a deterministic solution that receives weak gradients and negligible parameter updates. Given sufficient training steps, any model that averages over this data-facing model will converge to the same deterministic solution, thereby obviating it altogether. To allow for some useful weight averaging in this high-negentropy setting (while also allowing for an averaged model that is eventually approximately deterministic), we revolve parameters from the averaged model into the data-facing model at the end of each cascade stage. Practically, the average parameters from the previous stage of the cascade become the new data-facing parameters at the current stage, and a new averaged model is initialised (although it is also possible to decrease the weight of the running average instead of initialising a completely new averaged model). We use a large, constant learning rate of 0.05 during SWA steps.

\paragraph{Multiplier schedules and hyperparameters}
To train the parcellation model, we scheduled changes in the relative balance of loss multiplier hyperparameters over the course of training. We highlight the most critical aspect of this schedule, the entropy cascade, in the main text: we periodically increase the penalty on each vertex's parcel assignment distribution entropy to transition the parcellation solution from distributed and overlapping functional modes to compact and circumscribed deterministic parcels. Here, we discuss several additional scheduling decisions; the exact values of all multiplier hyperparameters and schedules are included in the associated code.

First, the loss terms that favour spatial and temporal separation of parcels---the dispersion and determinant terms---are initialised with large multipliers of 10 and .005 respectively in order to achieve rapid differentiation of parcels at the beginning of training. We anneal these multipliers to 0.5 and 0.0001 respectively before the 400th training step. Second, approximately following a practice established in previous parcellation efforts (e.g., \cite{schaefer_local-global_2018}), we progressively increase the compactness multiplier to collapse the spatial extent of parcels; we find that this increase works synergistically with the entropy cascade. Third, to reduce the chances of immediate fixation and convergence at the start of each entropy cascade step, we temporarily pump the multipliers for two terms that tend to compete with the entropy, the parcel equilibrium and second moment. These terms decay exponentially back to a baseline over the course of the cascade stage. Future work will more comprehensively evaluate the roles of these schedules in order to streamline the parcellation training regime.

\paragraph{Evaluation}
\begin{figure}
    \centering
    \includegraphics[width=\linewidth]{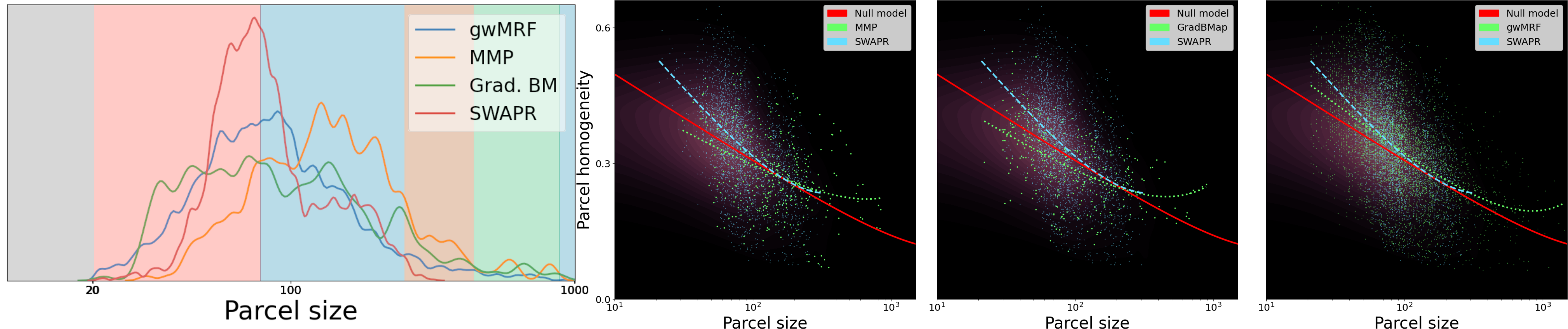}
    \caption{Parcellation evaluation. \textit{Left}, distributions of parcel sizes for different evaluated parcellations. Background colour indicates maximum scoring parcellation for that size according to log fit. \textit{Right}, Absolute homogeneity evaluation without adjustment for null.}
    \label{fig:atlas_suppl}
\end{figure}

The quality of the learned parcellation was evaluated using a measure of parcel homogeneity, operationalised as follows. For fairer comparison with reference parcellations, which are defined deterministically, we first took the arg max over vertex probability assignments to create a deterministic parcellation at each of the 5 evaluated parcel scales. (In practice, the final stages of the entropy cascade already yielded maximum assignment probabilities close to unity for nearly all vertices.) For each parcel at each scale, we then computed the pairwise Pearson correlation among all of that parcel's assigned vertex-wise time series for each BOLD image in the held-out test set. For each BOLD image, the parcel homogeneity was operationalised as the mean of these correlations. We then defined the overall homogeneity of that parcel as its mean homogeneity across all images in the test set. We note that the homogeneity scores we obtained here were low in relation to many previous reports; this might be attributable in part to the lack of spatial smoothing in our data. Additionally, of the reference parcellations, only the MMP parcellation \citep{glasser_multi-modal_2016} was defined using the HCP dataset; because of differences between datasets, acquisition protocols, and coordinate spaces in which the different parcellations were defined, comparison results should be interpreted with caution. As an additional note of caution, it is possible that the ``unseen'' data used for evaluation were not held out for the MMP parcellation. Furthermore, it should be noted that the reference gwMRF parcellation that we evaluated, like the dTSC/SWAPR parcellation, is in fact a set of parcellations defined across different spatial scales (all multiples of 100 parcels between 100 and 1000, inclusive).

Parcel homogeneity is related to parcel size---it is more likely for a smaller parcel to be more homogeneous. Accordingly, after computing homogeneity scores for each parcel, we also computed the best logarithmic least-squares fit of parcel size to parcel homogeneity. This fit was computed for all evaluated parcellations, including the null model. To generate the \textit{Relative Homogeneity} plots in the main text, we subtracted the null model's best fit from all other fits and homogeneity scores. As a reference, we also plot the analogous figures without this adjustment here (Figure \ref{fig:atlas_suppl}, \textit{Right}). Parcels with a size of under 20 vertices were excluded from evaluation. We also include KDE plots to facilitate visualisation of size distributions for different parcellations (Figure \ref{fig:atlas_suppl}, \textit{Left}). The parcel size distribution for our method reveals a tighter distribution than for other methods, likely due to a fairly stringent parcel equilibrium regularisation. In future work, we will explore the impact of relaxing this regularisation, particularly in consideration of the desirability for more variable parcel resolutions across the brain in certain applications. More flexible alternatives to a parcel equilibrium penalty, such as a unilateral L2 loss imposed on parcels whose total weight is less than some minimum, could also be explored.

\section{Denoising}

\begin{figure}
    \centering
    \includegraphics[width=\linewidth]{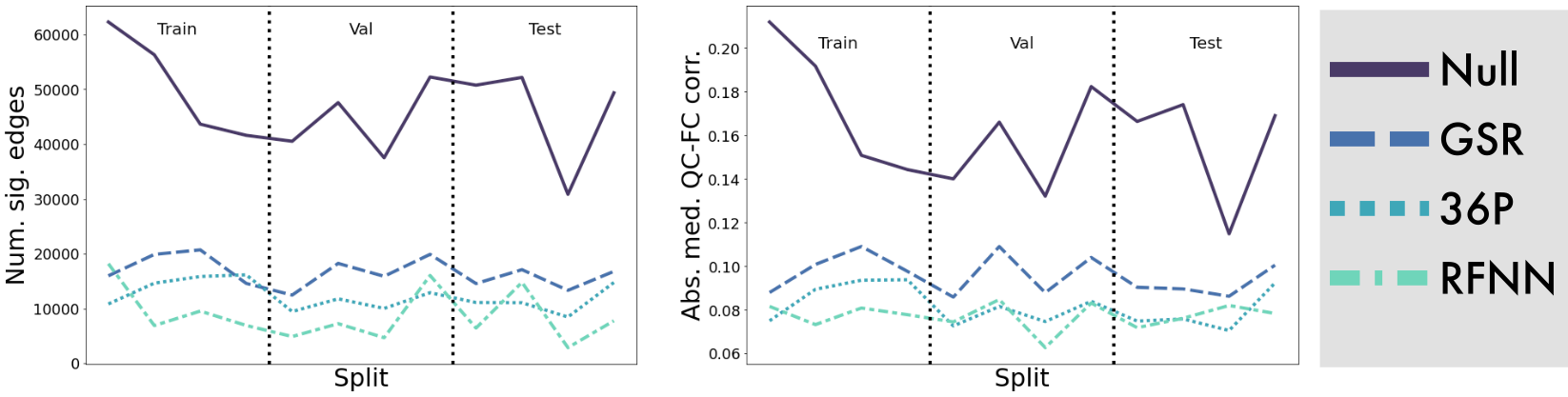}
    \caption{Denoising evaluation across all 12 data splits, each comprising around 350 images. The results shown in the main text are from the first test split. If the QC-FC benchmarks are taken at face value, the single-confound RFNN model outperforms other single-confound models with substantial consistency and performs competitively with the SOTA 36-confound model.}
    \label{fig:denoise_suppl}
\end{figure}

\paragraph{Dataset, preprocessing, and computational resources}
We again use the minimally preprocessed Human Connectome Project dataset detailed in the parcellation section above; however, we make a few adjustments. First, we use BOLD time series whose dimension has already been reduced via a mapping to parcels. We use the popular 400-region gwMRF parcellation introduced by \citet{schaefer_local-global_2018}, which is based on a Markov random field approach and recapitulates previously characterised boundaries of functional subsystems. Second, we use a different data split, training on 4 of the 12 random splits, selecting another 4 for validation, and setting the last 4 aside for evaluation. Due to computational limitations (because all data must fit into memory simultaneously for evaluation), we use a 351-image subset of the evaluation split for the main results. In the supplement, we report benchmark results using each of the 12 splits (Figure \ref{fig:denoise_suppl}).

Additionally, because the HCP dataset includes only gross motion estimates and ICA decompositions, we ran additional preprocessing steps to obtain a more complete complement of confound time series. For this additional preprocessing, we used stereotaxically embedded BOLD data volumes. First, we computed the mean time series across all brain voxels (the global signal, GS). We then computed the mean time series across voxels in binary masks indicating membership in white matter (WM) and cerebrospinal fluid (CSF) compartments; these masks were included with the HCP dataset. We also computed singular value decompositions of WM and CSF voxels, yielding orthogonal sets of principal component time series (aCompCor; \cite{behzadi_component_2007}); we included the first 10 from each compartment in the input to the RFNN model. Finally, we computed a standardised version of the DVARS, the standard deviation across all voxels in temporal difference images \citep{afyouni_insight_2018}. Following reports that head motion estimates can be contaminated by respiratory artefact \citep{fair_correction_2020}, we applied a notch filter to the gross estimates of head motion included with the HCP dataset. We then used these filtered estimates to compute the framewise displacement (FD) that we used as our main QC metric: the sum of the absolute values of the 6 filtered motion estimates \citep{power_spurious_2012}. We then computed an expansion of the 6 gross head motion estimates (translation and rotation along the x-, y-, and z- axes) and 3 compartment mean signals (GS, WM, and CSF); this expansion consisted of temporal derivatives obtained via backward differences, quadratic terms, and squares of derivatives. This expansion resulted in 36 terms, which together comprised the 36P model we used that has previously been demonstrated to give excellent performance among \textit{a priori} confound models \citep{satterthwaite_improved_2013,ciric_benchmarking_2017,parkes_evaluation_2018}.

In order to obtain relatively stable QC-FC correlations, the model was trained with a large batch size of 100 time series. After parcellation, data were further preprocessed as follows. First, BOLD data and \textit{a priori} confounds were normalised such that each time series had a mean of 0 and a variance of 1. Next, training and validation (but not evaluation) data were randomly windowed to select 500 contiguous time frames.

Because of incomplete acquisitions, data might be missing for some images. To address this, we created a missing data mask for each batch of time series. We then synthesised data to impute the missing time frames while minimally impacting actual time frames in downstream analyses. We used a hybrid approach for time series imputation: any missing epochs comprising 3 or fewer frames were imputed using a weighted average of the closest frames, while longer missing epochs were imputed using a periodographic approach. To elaborate, we created a basis of sine and cosine functions, selected the frames of each basis function that corresponded to seen epochs in the data, and fit each basis function to the data. We then used the estimated coefficients from all basis function fits to reconstruct unseen data as a linear combination of basis functions. This approach is inspired by a previous method introduced by \citet{power_methods_2014}, which in turn draws inspiration from the Lomb-Scargle periodogram \citep{lomb_least-squares_1976}. (We did not apply censoring to denoise the data for this experiment because, for this proof of concept, we were interested in optimising denoising model performance without censoring.) After this imputation step, time series were filtered to retain frequencies between 0.01 and 0.1 Hz using a brick-wall ideal filter applied multiplicatively in the frequency domain. The selected low-frequency band is relatively artefact-free \citep{satterthwaite_improved_2013} and corresponds reasonably with the frequency of the haemodynamic BOLD response. Filter weights were frozen and not configured to be learnable. To prevent reintroduction of previously orthogonalised signals during the denoising step \citep{hallquist_nuisance_2013}, identical imputation and filter transformations were applied to BOLD and confound time series. We again used several hours of compute time on a single RTX 2080 Ti GPU on a machine running Ubuntu Linux to train the RFNN model.

\paragraph{RFNN architecture, denoising, and loss}
The model we used for denoising is a minimal, shallow neural network with a single hidden convolutional layer. The hidden layer takes one channel of dimension $c \times t$---the $c = 57$ \textit{a priori} confound time series---as its input. It has a total $f = 5$ learnable filters, each of dimension $1 \times 9$. With appropriate padding and a stride of 1, the output of the convolutional layer is thus a set of $f$ $c \times t$ time series, which are the input time series convolved with each of the $f$ ``response functions''. These are passed through a thresholding leaky ReLU nonlinearity (subtracting a bias term, applying the nonlinearity, and adding back the bias term) before they are concatenated together with the input confound time series. The result is a $(f + 1)c \times t$ model matrix. The output layer of the RFNN linearly maps the $(f + 1)c$ confound time series to a single model confound.

To obtain a denoised connectome, we compute the covariance of the parcellated BOLD time series conditioned on this confound as defined in the main text. Weights of any imputed observations were set to 0, so denoised connectomes were based only on actual, seen data. We then normalise the conditional covariance to a Pearson correlation before computing the loss function. This symmetric correlation matrix is our estimate of the functional connectome. As described in the main text, the loss function (QC-FC) is a ``second-order'' correlation computed across the batch dimension between the correlation that represents each connectome edge and a measure of artefact, in this case the mean filtered framewise displacement across all time frames. Because an edge with an inverse relationship to motion is just as undesirable as one with a direct relationship, the loss is operationalised as the mean of the absolute value of QC-FC correlations. We train the RFNN model for 200 epochs using stochastic gradient descent (SGD), sampling 10 batches randomly per epoch. We also create a null model by randomly initialising a RFNN model without training it.

\paragraph{Evaluation}

We evaluate model performance using standard benchmarks derived from edge-wise QC-FC correlations \citep{power_methods_2014}. Benchmarks are computed on a held-out dataset of 351 BOLD time series. Benchmarks include the number of edges for which a significant relationship with motion is detected across subjects and the median of the absolute value of all QC-FC correlations.

Although it is not a focus of the current proof of concept, QC-FC correlations also tend to be distance-dependent: motion artefactually inflates estimates of short-range connections more than it does estimates of long-range connections. Furthermore, methods that remove the global signal are especially effective at removing widespread artefact that acts over longer ranges, so this distance-dependence is apparently exacerbated after GSR (e.g., \cite{satterthwaite_improved_2013,power_methods_2014}). We report the ``third-order'' correlation between QC-FC correlations and inter-node separation in Supplemental Table \ref{tab:distdep}.

\begin{table}[]
    \centering
    \begin{tabular}{l r}
        \toprule
        Model & Dist. Dep. \\
        \midrule
        Null & -0.083395 \\
        GSR & -0.121429 \\
        36P & -0.138829 \\
        RFNN & -0.117902 \\
        \bottomrule
    \end{tabular}
    \caption{Distance dependence of QC-FC correlations (lower absolute value is better).}
    \label{tab:distdep}
\end{table}

There is arguably some circularity in using QC-FC correlations as the outcome measure after directly training our model to minimise the mean of absolute QC-FC correlations. We leave to future work any investigation of the limitations of QC-FC measures and a more complete characterisation of the reasons that the RFNN was able to ostensibly outperform or compete with SOTA methods on these benchmarks without concomitant zero-centring of connectome edges. In the interim, we reiterate our tentative recommendation from the main text against use of this method until a more thorough investigation is conducted.

\section{Covariance}

\paragraph{Time-by-time covariance (subnetwork detection)}
We again used the HCP dataset for the time-by-time covariance clustering experiment. Of our 12 HCP dataset splits, we selected 4 for training and 4 for evaluation. Vertex-wise BOLD data were mapped onto 400 parcel-wise time series using the 400-parcel dTSC/SWAPR parcellation that we trained in our first experiment. Because of the hemispheric tether regularisation that we used when learning the parcellation, we were able to impose a soft inter-hemispheric symmetry constraint on the subnetwork detection problem.

We used a batch size of 20 and selected a random window of 800 time frames from each parcellated time series when sampling a batch. Further preprocessing steps included imputation, filtering, and denoising, each implemented as described in \textit{Denoising} above. Denoising was performed using a 36-confound model with demonstrated efficacy at removing structured artefact from fMRI data \citep{satterthwaite_improved_2013,ciric_benchmarking_2017,parkes_evaluation_2018}. The model was trained for 600 epochs, each of which consisted of 15 training steps, using SGD. Training required several hours on a single RTX 2080 Ti GPU on a machine running Ubuntu Linux.

The objective of the clustering is to learn a $c \times p$ covariance parameter $\Theta$: an assignment of $p$ parcels to $c$ clusters, which we interpret as subnetworks of the brain. For this proof of concept, we selected as our learning objective the minimal nontrivial clustering into $c=2$ subnetworks. During each forward pass through the covariance layer, a softmax mapped the 2-dimensional cluster assignment of each parcel onto the set of Bernoulli distributions. Next, we computed the two time-by-time covariances parameterised by the matrices $\mathbf{\Theta}$ obtained by embedding the elements of each row of $\Theta$ along the main diagonal. These time-by-time covariances can be interpreted as dynamic profiles of the two detected subnetworks. Finally, the covariances were normalised to time-by-time Pearson correlations.

To perform the clustering, we used a loss function consisting of 5 terms. First, a dispersion term, equal to the negative L2 distance between the vectorised upper triangles of time-by-time matrices, promoted separation between dynamic profiles. To favour structured dynamics with larger correlations, we also imposed a symmetric L2 term that penalised distance of each entry in the correlation matrix from either -1 or 1. Entropy and equilibrium terms were used as in the parcellation experiment to promote eventual deterministic assignment to each of 2 approximately equal-sized subnetworks. As in the parcellation experiment, we started with a weak entropy term that we increased at the 500th step. The final term was a Jensen-Shannon divergence penalty placed on the distance between a parcel's subnetwork assignment distribution, and the assignment distribution of that parcel's analogue in the opposite hemisphere. This penalty resulted in a relatively symmetric subnetwork assignment.

Because of the substantial autocorrelation between temporally proximal BOLD frames, which is further inflated by the temporal filter that we apply, each time-by-time correlation matrix typically features large values near its main diagonal. To downweight the importance of these autocorrelations in the clustering, we took the Hadamard product of each time-by-time correlation matrix with a Toeplitz-structured exponential discounting matrix. Specifically, before it was passed to the loss function, each entry of the correlation matrix was scaled by the factor $1 - e^{-\lambda t}$, where $t$ denotes its offset from the main diagonal and $\lambda = 0.1$ is a discount hyperparameter. To smooth training and improve clustering repeatability, we also used stochastic weight averaging \citep{izmailov_averaging_2019}. We tuned model and training hyperparameters by repeating the analysis on our training set until we found hyperparameters that yielded a repeatable solution.

We then assessed the out-of-sample replicability of the learned subnetwork structure by training the model on a held-out evaluation set and assessing the convergence between the results across samples. Although there were subtle differences, the subnetwork assignments detected in the two samples were approximately the same after alignment (mean JS divergence over 400 parcels: 0.0052; 3 parcels' maximum assignments differed per hemisphere). In the main text, we show maps of each parcel's maximum subnetwork assignment, together with examples of dynamic profiles for two subjects from the training set. The dynamic profiles are windowed time-by-time correlation matrices parameterised by subnetwork assignments.

\begin{figure}
    \centering
    \includegraphics[width=\linewidth]{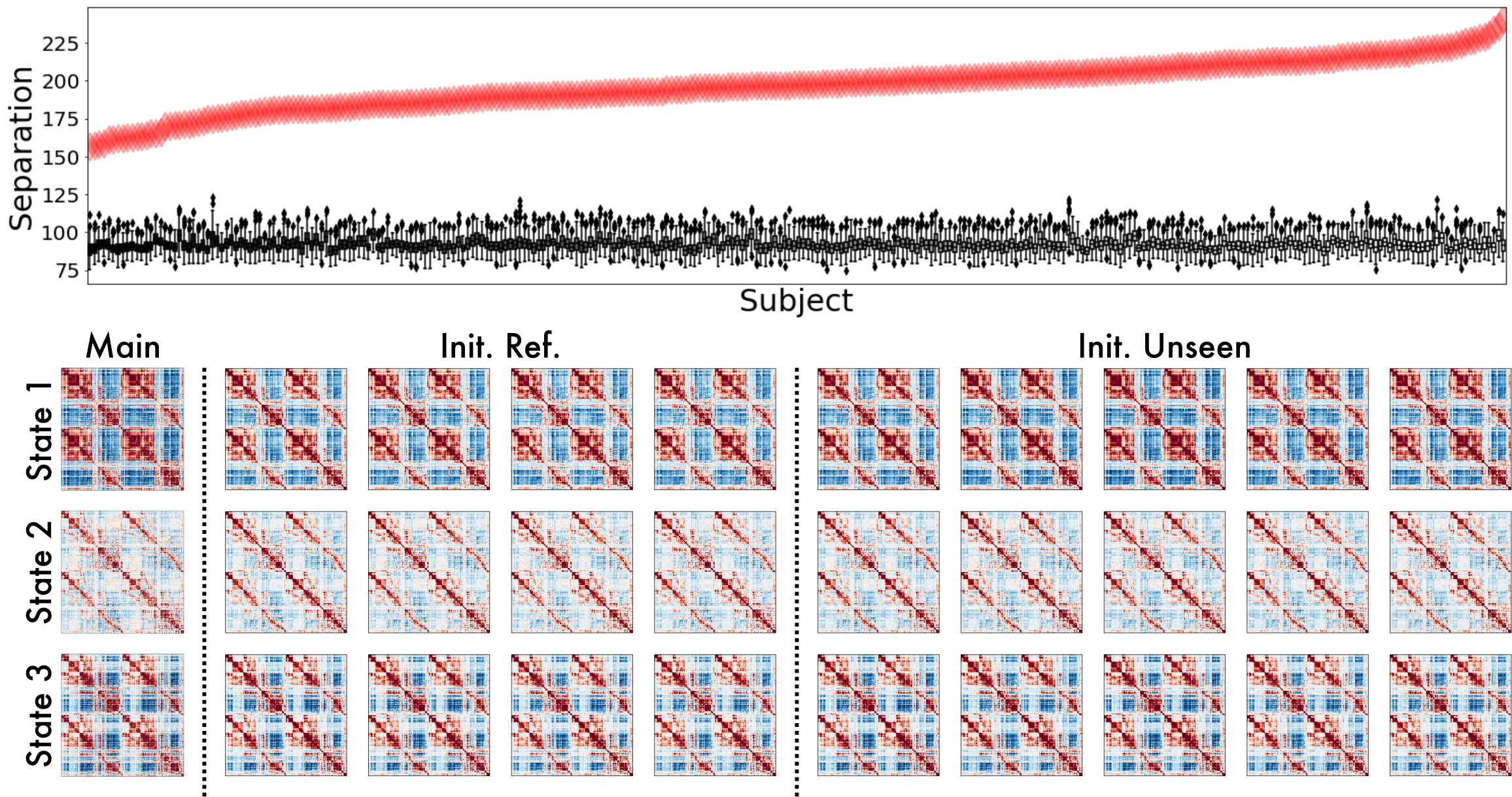}
    \caption{\textit{Top}, Compared with random symmetric subnetwork assignments, the subnetworks detected by clustering consistently achieved greater separation of time-by-time dynamic covariance matrices. \textit{Bottom}, The detected states were relatively consistent across both subsamples used for the initialisation and entirely unseen data.}
    \label{fig:cov_suppl}
\end{figure}

Here, we also run a further evaluation of the subnetwork detection (Figure \ref{fig:cov_suppl}, \textit{Top}). We used the subnetworks detected in the training set as a reference, and we created 200 null subnetwork models by randomly assigning each hemisphere's parcels to subnetworks while preserving the number of parcels per subnetwork and inter-hemispheric symmetry. We selected 300 subjects from the evaluation set and computed for each subject the L2 distance between the dynamic profiles of the learned subnetworks, as well as the L2 distance between the dynamic profiles of each null model's subnetworks. Figure \ref{fig:cov_suppl} compares the separation between dynamic profiles for the learned subnetworks (red) against null distributions (black) for each subject. In all cases, we find that the learned subnetwork assignments yield a superior separation to all null models.

\paragraph{Regional covariance (state detection)}
For the state detection experiment, we used small subsamples of HCP data. Each of the 12 data splits we created was further divided randomly into 5 shards, each consisting of approximately 50-80 resting-state BOLD time series. The results in the main text are for one shard with 59 time series; we found this to be sufficient for stable state detection. We show these results together with 9 replicates in Figure \ref{fig:cov_suppl}.

Preprocessing followed a standard functional connectivity pipeline. The first stage was parcellation using the 400-parcel gwMRF atlas \citep{schaefer_local-global_2018}, which we selected because its parcels are ordered to follow the brain's large-scale community structure and would thus be conducive to visualisation of differences between states. The 400 parcel-wise time series were then processed through imputation, filtering, and denoising as described in the \textit{Denoising} section. We again denoised the time series using the proven 36-confound model, which has demonstrated efficacy in removing structured artefact due to motion and respiration from BOLD time series \citep{satterthwaite_improved_2013,ciric_benchmarking_2017,parkes_evaluation_2018}. The state detector model was trained for 50 epochs. Training required several minutes on a 12-core first-generation Threadripper CPU on a machine running Ubuntu Linux.

We used $k$-means clustering to initialise $k = 3$ learnable state \textit{templates}. For this proof of concept, we select $k = 3$ because it is the smallest number of clusters that does not produce only simple high-connectivity and low-connectivity states. The templates are initialised by first computing sliding-window correlations (as in \cite{allen_tracking_2014}) with a window size of 50 and a sliding step size of 25 for each of 1180 total images that constitute 4 splits of the HCP dataset. All windows are submitted as observations to the $k$-means clustering algorithm. We use the L2 distance to cluster in this proof of concept because it is compatible with various common implementations of $k$-means; however, a cosine distance might better capture differences in connectivity configurations that are not simply due to connection magnitude. Each of the $k$-means states that we detected using $k$-means clustering approximately corresponded with a single state that we later detected using our differentiable approach. The $k$-means initialisations additionally formed a baseline for model evaluation.

The learnable parameters of the clustering model include the templates thus initialised and instance-specific $c \times t$ covariance parameters $\Theta$, where we chose $c = k = 3$. Each of the $c$ parameter vectors of length $t$ can be interpreted as the time course of a state; we interpret a value close to 0 as the absence of the state, and a value close to 1 as the presence of the state at some time point. We require an \textit{instance-specific} parameterisation for each state because the presence or absence of a state is likely to occur at different times for each subject under the unconstrained conditions of the resting state. The use of instance-specific parameterisations is not handled natively by optimisers included with many existing deep learning software libraries. This is likely in part because of the ambiguity over how to handle buffers associated with instance-specific parameters (such as parameter-specific moments). For this proof of concept, we take the most direct approach of maintaining a separate buffer for each instance. We thus implemented new optimiser classes to handle parameter groups that exist ephemerally in memory. (This applies to the instance-specific parameters that we use here, but could also be extended to parameters specific to subgroups of the dataset--for instance, subjects with several runs.)

To train the state detection model, we used an instance-specific extension of SGD with a loss function consisting of 5 terms. First, an L2 penalty was applied to the backwards temporal difference of the instance-specific state time courses $\Theta$ in order to promote state persistence and increase the evidence required for the model to predict a transition between states. Second, an equilibrium loss was imposed to ensure that all states were represented in each instance. Third, a dispersion penalty was used to promote separation of each instance's detected states. The two remaining loss terms used the learnable templates to align detected states across instances. One term, another dispersion penalty, was used to promote separation of templates, and the final term was a penalty on the L2 distance between each instance's three detected states and the three templates. This final term encouraged the learnable templates to function as population-level representations of the states detected across instances. We did not use an entropy penalty for this clustering problem.

The main text figure shows the learned parameters $\Theta$ and the learned templates, and compares the L2 distance between the states detected by the $k$-means initialisation and the averages (across instances) of analogous states detected by the differentiable approach. We also repeated the state detection experiment in 4 additional subsamples of the HCP dataset, each comprising between 50 and 80 instances, which were used in computing the $k$-means initialisation, and 5 similar subsamples that were not used in computing the initialisation. We qualitatively observed consistent detection of three analogous states across all subsamples (Figure \ref{fig:cov_suppl}, \textit{Bottom}).

\section{Community detection}

\paragraph{Dataset and preprocessing}
We use the Midnight Scan Club dataset (MSC) for the community detection analysis. The MSC dataset includes a total 10 subjects each densely scanned over 10 sessions, both at rest and performing a number of directed cognitive tasks \citep{gordon_precision_2017}. MSC data are openly available and were acquired in accordance with the guidelines of the host institution's IRB. In this proof of concept, we limit our analysis to resting-state data. We use the first three scans from each subject for the community detection analysis.

Data were parcellated using the 400-parcel gwMRF parcellation \citep{schaefer_local-global_2018}, filtered, and denoised using a 36-confound model with demonstrated efficacy at removing structured artefact associated with motion and respiration from BOLD data \citep{satterthwaite_improved_2013,ciric_benchmarking_2017,parkes_evaluation_2018}. The denoised, parcellated BOLD time series were inputs to the community detection model.

The learnable parameters of the model were (i) the $p \times c$ community affiliation matrix $\mathbf{C}$ of $p$ parcels to $c$ communities and (ii) the instance-specific $c \times t$ time courses $\Theta$ of each community. The community affiliation of each parcel was mapped through a softmax function that ensured it was a valid probability distribution, and each community time course was mapped through a sigmoid that constrained it to $(0, 1)$. Each community time course could thus be interpreted as the extent to which the corresponding community was modularised during each time frame of the scan. Each of these community time courses thereafter parameterised a separate covariance matrix.

\paragraph{Model details}
We thus computed $c = 10$ parameterised $400 \times 400$ covariance matrices $\mathbf{A}_{\mathcal{C}_i}$ and 1 unparameterised (standard or ``time-averaged'') $400 \times 400$ covariance matrix $\mathbf{A}$ among the 400 parcel-wise BOLD time series. (All covariance matrices were normalised to Pearson correlations for the purpose of further analysis.) Next, we used the community affiliation $\mathbf{C}$ to estimate the relaxed Girvan-Newman modularity \citep{newman_finding_2004} on the unparameterised covariance matrix $\mathbf{A}$ as described in the main text. We then considered each column of $\mathbf{C}$ separately. Each column $\mathbf{C}_i$ of $\mathbf{C}$ corresponds to a single community $\mathcal{C}_i$, and the entries of the rank-one outer product $\mathbf{C}_i \mathbf{C}_i^\intercal$ indicate the extent to which each edge of the connectome graph connects two vertices in $\mathcal{C}_i$. We used each of these low-rank outer products as coaffiliation matrices $\mathbf{H}_{\mathcal{C}_i}$ to compute the relaxed Girvan-Newman modularity on the corresponding parameterised covariance $\mathbf{A}_{\mathcal{C}_i}$.

We thus sought to optimise the $c + 1$ modularities by jointly learning the affiliation of each parcel to a community and the extent to which the community was modularised during each time frame. To achieve this objective, we trained the model using a loss function with four main terms. Two multipliers controlled the balance between static communities parameterised only by $\mathbf{C}$ and dynamic communities parameterised by $\mathbf{C}$ and $\Theta$. The remaining two multipliers regularised the learned parameters: one modulated an L2 penalty on the minimum distance between each entry in $\Theta$ and either 0 or 1, thereby steering the model to make a binary decision for each time point as to whether a community was modularised or not, and the other modulated a backward difference L2 penalty on $\Theta$ that promoted persistence of modularisation and demodularisation states for each community. The model was trained for 500 epochs using the Adam optimiser \citep{kingma_adam_2017}. Training required under 1 hour on a 12-core, 24-thread CPU.

\paragraph{Evaluation}
To evaluate the validity of the learned community time courses $\Theta$, we created, for each instance, a distribution of 500 null community time courses by randomly shifting that instance's learned time courses. Time frames that the shift displaced over the end of each time course were wrapped back to its start. We then computed, for each of the null covariances $\widetilde{\mathbf{A}}_{\mathcal{C}_i}$ parameterised by a randomly shifted time course, the corresponding modularity. We created the main text plot of learned and null modularities for each community by subtracting each of the 500 null modularities computed for each instance from the learned modularity for that instance. We generally found that the learned modularity outperformed random-shift nulls. (We remark that small random shifts, and even random shifts of zero, are likely to occur because 500 random shifts are selected for each subject, and each time course is 814 frames long. Due to factors including the persistence imposed on each community's modularisation state and the substantial autocorrelation of filtered BOLD time series, these small random shifts will likely have modularity very close to that of the learned model.)

\printbibliography[heading=subbibliography]

\end{refsection}

\end{document}